\documentclass[namedreferences]{SolarPhysics}
\usepackage[optionalrh]{spr-sola-addons} 
\usepackage{graphicx}        
\usepackage{color}           
\usepackage{url}             

\newcommand{\text}[1]{\mbox{#1}}

\newcommand{\aap}{    {\it Astron. Astrophys.}}

\newcommand{\aj}{     {\it Astron. J.}}
\newcommand{\apj}{    {\it Astrophys. J.}}

\newcommand{\solphys}{{\it Solar Phys.}}

\begin{document}

\begin{article}

\begin{opening}

\title{Dust detection by the wave instrument on STEREO:\\ nanoparticles picked up by the solar wind?
}

\author{N.~\surname{Meyer-Vernet}$^{1}$\sep
        M.~\surname{Maksimovic}$^{1}$\sep
        A.~\surname{Czechowski}$^{2}$\sep
        I.~\surname{Mann}$^{3}$\sep
        I.~\surname{Zouganelis}$^{1, *}$\sep
        K.~\surname{Goetz}$^{4}$\sep
        M.~L.~\surname{ Kaiser}$^{5}$\sep
        O.~C.~\surname{St. Cyr}$^{5}$\sep
        J.-L.~\surname{Bougeret}$^{1}$\sep
        S.~D.~\surname{Bale}$^{6}$           
                   }
\runningauthor{Meyer-Vernet et al.}
\runningtitle{Solar wind nanoparticles}

   \institute{$^{1}$ LESIA, Observatoire de Paris, CNRS, UPMC, Universit\'{e} Paris Diderot; 5 Place Jules Janssen, 92190 Meudon, France
                     email: \url{nicole.meyer@obspm.fr}\\ email: \url{milan.maksimovic@obspm.fr}\\ 
              $^{2}$ Space Research Centre, Polish Academy of Sciences, Bartycka 18 A PL-00-716, Warsaw, Poland 
                    \\ email: \url{ace@cbk.waw.pl}
 \\
$^{3}$ School of Science and Engineering, Kindai University Kowakae 3-4-1, Osaka, 577-8502,
Japan   \\                   email: \url{mann@kindai.ac.jp}
 \\
$^{4}$ School of Physics and Astronomy, University of Minnesota, Minneapolis, USA
                   email: \url{goetz@waves.space.umn.edu}  
\\
$^{5}$ NASA-GSFC, Greenbelt, MD 20771, USA \\
                   email: \url{michael.kaiser@nasa.gov} \\
$^{6}$ Space Sciences Laboratory, University of California, Berkeley, USA \\
                     email: \url{bale@ssl.berkeley.edu} \\
$^{*}$ Now at: LPP, UPMC, EcolePolytechnique, CNRS, 4 av. de Neptune, 94107 Saint-Maur-des-Foss\'{e}s, France  \\
email:  \url{yannis.zouganelis@lpp.polytechnique.fr}
   }

\begin{abstract}
The STEREO wave instrument   (S/WAVES)   has detected a very large number of intense voltage pulses. We suggest that these events are produced by impact ionisation of nanoparticles striking the spacecraft at a velocity of the order of magnitude of the solar wind speed. Nanoparticles, which are half-way between micron-sized dust and atomic ions, have such a large charge-to-mass ratio that the electric field induced by the solar wind  magnetic field  accelerates them very efficiently. Since the voltage  produced by dust impacts increases very fast with speed, such nanoparticles produce signals as high as do much larger grains of smaller speeds. The flux of  10-nm radius grains inferred in this way is compatible with the interplanetary dust flux model. The present results may represent the first detection of fast nanoparticles in interplanetary space near  Earth orbit.  
\end{abstract}
\end{opening}

\section{Introduction}
     \label{S-Introduction} 

The  wave instrument \cite{bou08} on the STEREO spacecraft  was designed to study solar radio emissions and solar wind plasma waves. It also detects frequent wave bursts that cannot be explained with the known solar wind plasma emissions  (Figure~1). The spectral shapes of these events are similar to those first recognised as dust impacts in Saturn's G ring on Voyager (Aubier \textit{et al.}, 1983; Gurnett  \textit{et al.}, 1983)  and then observed in many other environments  (see a review in Meyer-Vernet 2001), which have been mainly interpreted by impacts of micron-sized dust moving at Keplerian speeds.  Basically, the energy dissipated upon impact vaporises and partly ionises the grain as well as a part of the target's material  \cite{dra74}. This produces an expanding plasma cloud whose residual charge induces an electric  pulse which is detected by the antennas - a process akin to that used by grain mass spectrometers in space \cite{gol89}.

If these STEREO wave burst observations in the solar wind at 1 AU were produced by impacts of micron-sized grains \cite{kai07}, the observed impact rate and signal amplitude would imply a flux exceeding that suggested by the interplanetary dust model by about four orders of magnitude. On the other
hand, the observed impact rate is consistent with particles of much smaller size - in the nano-meter range,  which are much more numerous since the interplanetary flux is expected to increase fast as mass decreases (Figure~5).

We hence study whether impacts of nanoparticles on the spacecraft can produce such wave signals. That this is indeed the case is suggested by a recent analysis of the wave data acquired on the spacecraft Cassini in the solar wind during the Jupiter fly-by, which are consistent with impacts of high speed nanoparticles detected simultaneously by the on-board dust analyzer \cite{mey08}. The latter  nanoparticles move fast because they have a very large charge-to-mass ratio, so that the jovian corotational electric field accelerates them very efficiently. And since the charge released upon impact on
the spacecraft increases very fast with speed, they produce wave pulses as high as do much larger grains moving more slowly.

We here propose that the numerous voltage pulses observed on STEREO are also produced by impacts of high speed nanoparticles. The micrometer-sized dust mentioned above  is mainly affected by the gravitational force and moving on roughly Keplerian orbits, so that the impact speeds onto the spacecraft are of the order of a few 10 km s$^{-1}$. The smaller dust particles in the 0.1 to 1 micrometers range - the so-called $\beta $ meteoroids, are affected by the solar radiation pressure and ejected  at speeds of the order of several 10 km s$^{-1}$  \cite{weh99}. On the other hand, for particles of size below about 10 nm the Lorentz force is dominant, and the gyrofrequency is much greater  than the orbital frequency. In this case the final velocity of ejected particles is of the order of magnitude of the solar wind speed  \cite{man07}.

The acceleration process is similar to that of freshly produced ions in the solar wind  \cite{luh03}, which produces pickup ions from interstellar gas  \cite{mob85} and from dust interaction with the  solar wind   \cite{gei95}. The nano particles are produced by collisional fragmentation of the larger dust particles and their production rate increases with increasing dust number density and dust relative velocities toward small distances from the Sun.

In the following we describe these  STEREO wave observations and estimate the signal expected from dust impacts. We then estimate the speed expected for nano particles and derive their charge production upon impact onto a target. We finally deduce an average flux of nano particles from the observed  voltage spectral density and show that this flux is close to the order of magnitude predicted by the model distribution of interplanetary dust at 1 AU. Unless otherwise indicated, units are SI.

\section{Wave observations}
\label{S-wave}  

The  spacecraft STEREO A and B travel close to Earth's orbit, respectively ahead and behind, at angular distances increasing by $22^{\circ }$/year.  The S/WAVES instrument measures separately the waveform  and the power spectral density of the voltage \cite{bou08} at the ports of three 6-m antenna booms   \cite{bal08}.

 \begin{figure}    
   \centerline{\includegraphics[width=0.9\textwidth,clip=]{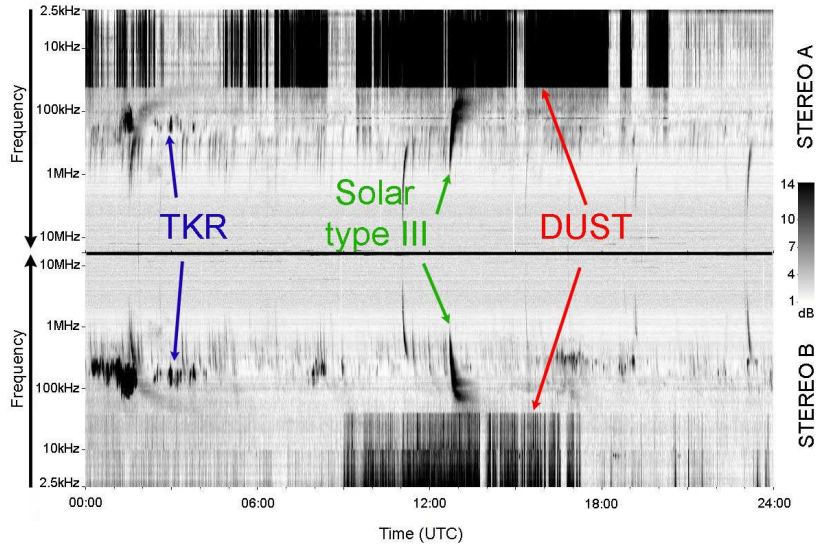}               }              
              \caption{Typical spectrograms, displayed as frequency versus time with relative intensity above the background scaled in grey, acquired by the WAVES low-frequency receivers with the X-Y dipole on STEREO A and B (then separated by  $2^{\circ }$) showing  voltage pulses, solar type III bursts and terrestrial kilometric radiation (TKR). The discontinuity between the 3 bands of the receiver for the transient pulses is produced by the different integration times (see the text).                    
   \label{fig1}    }
   \end{figure}

The antennas have an asymmetric configuration, with the three mutually orthogonal booms  (X, Y, Z on  Figure~2) mounted on a corner of the anti-sunward face of the spacecraft, at about $125^{\circ }$ to the Sun-spacecraft line. Thus a large part of the Z boom, which extends along the anti-sunward face,  lies in the wake of the spacecraft, whereas the X and Y booms extend respectively  close to and far from the spacecraft. This configuration also produces an asymmetry between both spacecraft, since the booms are mounted close to respectively the  orbital ram and rear face for STEREO A and B. Hence, since orbital motion tilts the wake axis with respect to the sun-spacecraft line, the parts of the X and Y booms closest to the spacecraft are  respectively outside and within the wake for STEREO A and B. As a result,  there is only one boom that is both close to the spacecraft and fully outside the wake: the X-boom, and this occurs only on Stereo A; this is also the boom that shows the largest wave pulses.

 \begin{figure}    
   \centerline{\includegraphics[width=0.6\textwidth,clip=]{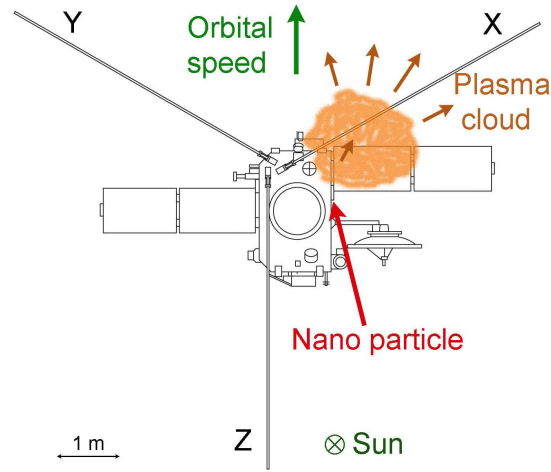}
              }
              \caption{View of the STEREO A spacecraft  from an observer looking towards the Sun,  showing  the three antenna booms X, Y, Z, and a sketch of the  expanding plasma cloud produced by the impact on the spacecraft of a nano particle moving in the prograde sense at several 100 km s$^{-1}$. STEREO B is similar, with a rotation by $180^{\circ }$  around the Sun/spacecraft line. The plasma cloud is drawn with the radius $R_2$ defined in Section 3, with the scale shown at the bottom-left. 
    \label{fig2}  }
   \end{figure}

Figure~1  shows  typical  spectrograms exhibiting voltage pulses, and Figure~3  shows an example of the power spectral density recorded on the low-frequency receiver with the X-Y  dipole during periods of voltage pulses  on STEREO A. The insert contains a typical waveform recorded by the time-domain sampler of the instrument; note that simulations have shown that the observed undershoot of the pulse should be  produced by instrumental filtering.

The low-frequency receiver consists of three 2-octave frequency bands which are sequentially connected to the antennas, with  acquisition times inversely proportional to the mid-band frequency. The observed power spectrum has a $f^{-4}$ shape which flattens at lower frequencies - a generic behaviour for pulses of rise time greater than $1/2\pi f$, as observed in previous dust detections by wave instruments (Meyer-Vernet, 2001,  and references therein).  The amplitude often exhibits a discontinuity at the low end of the middle band, whose response is affected by the non-stationarity of the dust signal because of the short acquisition time. The acquisition time of the higher band  is still shorter, so that it is not much affected by dust impacts and thus shows the ubiquitous plasma quasi-thermal noise in the solar wind, plus the  $f^{-2}$ shot noise due to photoemission and ambient electron impacts \cite{mey89}.

 \begin{figure}    
   \centerline{\includegraphics[width=0.6\textwidth,clip=]{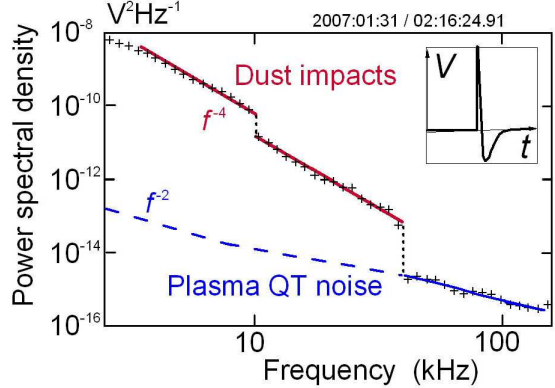}
              }
              \caption{Example of power spectrum with a typical individual wave-form inserted. The three frequency bands (of limits shown by dotted vertical  lines) are sequentially connected with an acquisition time inversely proportional to  frequency. The spectra in the two lower bands are produced by dust impacts; the higher band detects the ubiquitous plasma quasi-thermal and shot noise (blue, Meyer-Vernet  and Perche, 1989).
  \protect   \label{fig3}  }
   \end{figure}

For pulses of rate $N$, maximum amplitude  $\delta V$, and rise time $\tau $,    the  theoretical power spectrum  can be shown to be
\begin{eqnarray}
V_f^2 \simeq 2 <N  \delta V ^2 \omega ^{-2}(1+\omega ^2 \tau ^2)^{-1} >   \label{Vf2}
\end{eqnarray}
at frequencies $f= \omega /2  \pi $ much greater than the pulses' inverse decay time  \cite{mey85}, where the angular brackets stand for averaging over the pulses detected during the acquisition time. Comparison with (\ref{Vf2}) of the observed spectral shape in the lower band (which has the greatest acquisition time)   shows that  on average during these events on STEREO A, we have
\begin{eqnarray}
<  N \delta V^2/\tau ^2 > ^{1/2} \sim 2\times 10^4 \text{ Volts}\text{ s}^{-3/2}  \label{meas}
\end{eqnarray}
at the preamplifier input, taking into account the receiver gain $\Gamma \simeq 0.5$, where the averaging involves a large number of pulses, and $\tau \sim 4 \times 10^{-5}$ s.

These observations are summarised in Figure~4  which shows the  squared voltage observed with the  X-Y dipole on STEREO A (normalised to $f_{\rm{kHz}}^{-4}$ and averaged in the lower band) as a function of ecliptic longitude in 2007.  Further analysis shows that this signal is mainly due to the X boom (the Y boom signal being generally much smaller), and that the power on the Z boom is also generally smaller by at least one order of magnitude. These wave bursts are observed much less frequently on STEREO B, and with a smaller power.

 \begin{figure}    
   \centerline{\includegraphics[width=0.7\textwidth,clip=]{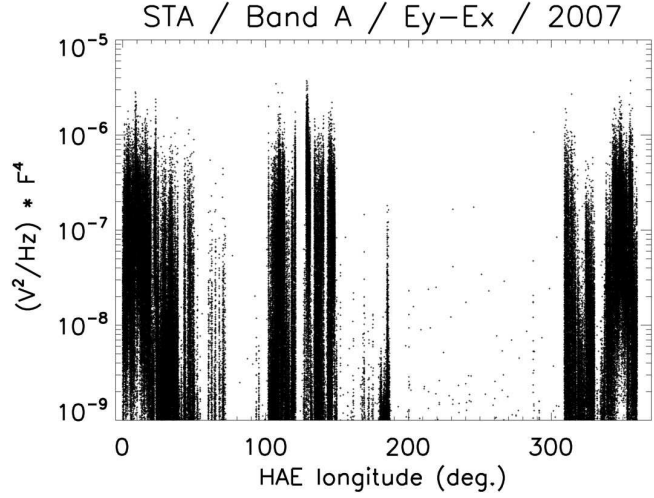}
              }
              \caption{Average power observed by  the STEREO/WAVES low frequency receiver ($V_f^2$ in V$^2$Hz$^{-1}$, normalised to $f_{\rm{kHz}}^4$ and integrated in the lower band) on STEREO A as a function of ecliptic longitude during the  whole  2007 orbit.
   \label{fig4}              }
   \end{figure}

\section{Electric signal of dust impacts}

Consider a dust grain which impacts the spacecraft, vaporises and ionises, producing an expanding plasma cloud. For a preliminary order of magnitude estimate, we model this cloud as made of $Q/e$ elementary charges of each sign at temperature $T$,  moving and expanding at velocity $v_E$; here $T$ is a kinetic temperature, not implying thermal equilibrium. Several scales enter the problem: the cloud's radius $R\simeq v_E t$  after time $t$, so that the plasma density is $n \simeq  3Q/(4 \pi R^3 e) \propto t^{-3}$, the cloud's proper Debye length
\begin{eqnarray}
L_{D} \simeq (4 \pi \epsilon _0 k_B T R^3/3Qe)^{1/2} \propto t^{3/2}\text{ ,} \label{LD}
\end{eqnarray}
the spacecraft size $\sim 1-2 $ m (not including appendages), and the ambient Debye length, which depends on the impact site and geometry. The latter is $ \sim 10$ m for the unperturbed solar wind,  $ \sim$ 1 m for the sunward photoelectron sheath (proportional to the inverse square root of the photoelectron yield of the sunward face), and much larger  in the hot and depleted spacecraft wake, where the charge dynamics is governed by the large wake-induced negative potential.

At the beginning of the expansion, the plasma cloud is so dense that its proper Debye length $L_D$ 
 is smaller than its radius $R$, so that it remains quasi-neutral. When $L_D$ becomes greater than $R$, charge decoupling occurs, so that the cloud's electrons, of speed $ v_e \sim (k_BT/m_e)^{1/2}$, can leave. From (\ref{LD}), one sees that this takes place when the cloud's radius $R$  reaches the value
\begin{eqnarray}
R_1=  3Qe/4 \pi \epsilon _0 k_B T
\end{eqnarray}
after a time  $\tau _1 \sim R_1 /v_E$. The density in the cloud is then roughly equal to $n  \simeq 1.8 \pi ^2 (\epsilon _0 T_{\rm{eV}})^3/(eQ^2) $, where $ T_{\rm{eV}}$ is the cloud's temperature in eV.

In the present case, this cloud density $n$ is  generally still greater than the ambient density $n_a$, and   $R_1$ is smaller than the ambient Debye length (which results from $n>n_a$ since the cloud temperature $T<T_a$, the ambient  temperature), so that a voltage can be detected. This holds until the  density $n$ in the cloud has decreased to the ambient value $n_a$, which takes place when   the radius of the cloud is
\begin{eqnarray}
R_2 \sim (3Q/4 \pi e n_a)^{1/3}\text{ .}    \label{R2}
\end{eqnarray}
Since the spacecraft equilibrium potential is expected to be a few times the photoelectron energy, which is smaller than the solar wind electron energy,  the ambient density $n_a$ is close to the solar wind density (except in the sunward photoelectron sheath where it is greater, and in the wake of the spacecraft where it is smaller). The scale $R_2$ given in (\ref{R2}) gives the maximum radius of the plasma cloud, which is reached 
 at the time
\begin{eqnarray}
\tau_2 \sim R_2 /v_E\text{ .}    \label{tau2}
\end{eqnarray}

Since the surfaces are charged positively because the ejected photoelectrons dominate the charging process in the absence of dust impacts, the electrons are easily recollected by the spacecraft of capacitance $C$, producing a voltage pulse
\begin{eqnarray}
\delta V_1 \sim - Q/C\text{ .}    \label{V1}
\end{eqnarray}
It can be detected (with opposite sign) by a monopole antenna that measures the difference of potential between the antenna boom and the spacecraft.

For  monopole antennas which are long and not very close to the spacecraft, charge recollection by the spacecraft as described above  is generally the main process governing dust detection  \cite{mey96}. Since the signal is a variation in spacecraft potential, it produces similar voltages  on different (symmetrical) monopole antennas. On the  other hand, dipole antennas, which respond to the difference in voltage between  two booms, detect  spacecraft voltage pulses only via imbalances. This results in a detected voltage much smaller than $\delta V_1$. This effect was first observed by the wave experiments on the spacecraft Voyager, when the PRA instrument using monopole antennas detected much higher (and better calibrated) voltage pulses produced by dust impacts than the PWS instrument using dipoles \cite{mey96}. Unfortunately for dust detection, most wave experiments in space are using a (symmetrical) dipolar configuration.

The case of STEREO is more complex because  the antennas are very asymmetric, very short,  and very close to the spacecraft - being fully within its Debye sheath, in contrast to most interplanetary probes carrying wave instruments. Hence  the electric field   \cite{obe96} induced by the plasma cloud on  the antennas (and possibly the charge collection by them) is expected to play a major role, strongly dependent on the impact site and geometry. This is further complicated by the various appendages, including the large solar panels which contribute to a conspicuous wake.

The maximum cloud radius $R_2$ (shown to scale on Figure~\ref{fig2}) is comparable to the spacecraft size and not much smaller than the antenna length. Therefore, for an impact on the spacecraft at a distance from the antenna closer than about $R_2$, an appreciable part of the antenna - a fraction $R_2 /L$ of its length, may lie within the plasma cloud. In this case the amplitude of the  voltage pulse induced on an antenna  boom of length $L$  is
\begin{eqnarray}
\delta V_2 \sim (k_B T/e)R_2/L    \label{V2}
\end{eqnarray}
with a rise time in the order of magnitude given by (\ref{tau2}).  On the other hand, if the distance of the impact site $r \gg R_2$, the induced voltage $ \sim Q/(2 \pi \epsilon _0 L) \ln (L/r)$ is much smaller.

For the nanoparticles detected by STEREO, we have $\delta V_2 \gg \delta V_1$. Thus the voltage pulse produced on an antenna boom in the expanding plasma cloud is much larger than that produced on the spacecraft. Consequently, this boom will detect a pulse of large amplitude $\sim \delta V_2$, whereas the other monopoles detect a pulse of much smaller amplitude  $\sim \delta V_1$. Hence, we will use (\ref{V2})  for interpreting the pulses on the X antenna of STEREO A, which  are detected by the frequency receiver at the ports of  the X-Y dipole.

The signal produced in this way is thus expected to be much greater than that on other spacecraft carrying wave instruments as for example Wind, Ulysses, and Cassini. Their electric antennas are symmetrical, longer, and do not extend close to the spacecraft surface, so that they detect dust grains mainly via the pulses in spacecraft potential  (\ref{V1}), which are of much smaller amplitude than (\ref{V2}). Furthermore,  these are detected with a much reduced amplitude in dipole mode - the most frequent configuration. This was  confirmed by the recent observation  of nanoparticles near Jupiter by the Cassini wave experiment \cite{mey08}.

Finally, it should be noted that  the time for the cloud electrons to travel    $L_D$, i.e. the inverse of the cloud's plasma frequency, $\omega_p^{-1} = \tau _1 (R/R_1)^{3/2}(v_E/v_e) $ is smaller than $ \tau _1$ for $R\sim R_1$ since in general $v_e \gg v_E$. This raises the possibility of charge oscillation in the cloud  producing plasma waves, which may, however, be hampered by the fast temporal decrease of $\omega_p$.  The fast motion of the charges might also produce acoustic waves.   Note, too, that the various appendages may enable ejected debris to  reimpact,  increasing the pulse amplitude and rise time.

\section{Dust speed and released charge}

To estimate the  charge $Q$ released upon a grain impact on the spacecraft, we must calculate the grain speed.  Interplanetary dust grains are charged under the influence of photoemission, plasma impact, and secondary emission. Since photoemission is in general heavily dominant, the grains acquire a positive electric potential $\phi$ of several times the photoelectron energy.  A grain of radius $a$ and mass density  $\rho$ thus carries a charge $q \simeq 4 \pi \epsilon _0 a \phi$ and has a charge-to-mass ratio $q/m \simeq 3 \epsilon _0  \phi /a^2 \rho$ that increases fast as size decreases.  For $a\sim 10$ nm ($m\sim 10^{-20}$ kg for   $\rho \simeq 2.5 \times 10^3$ kg m$^{-3}$) this yields $ q/m \sim  10^{-5} e/m_p $, 
where $e$ and $m_p$ are respectively the electron charge and proton mass, and the nanoparticle potential  at 1 AU is $\phi \sim 5-10$ V  \cite{muk81}, which is below the limit for ion field emission and for electrostatic disruption for most materials  \cite{men01}.

With such a $q/m$ ratio, the Lorentz force is dominant and must be considered when calculating the dust dynamics. In order to be accelerated efficiently, the grains should  have  $q/m$ large enough that their Larmor frequency be much higher than the frequency of orbital motion. Assuming that the grains are initially released between 0.1 and 1 AU into Keplerian orbits, this can be seen to be true for $q/m\geq 10^{-5}e/m_p$, so that 10 nm or smaller grains reach a velocity of the order of solar wind velocity.

To be more quantitative, we estimated the expected velocity distributions of these particles at 1 
AU from a sample of numerically calculated trajectories of grains. A 
 simple model of the solar wind (constant radial velocity) and of its magnetic field (Parker spiral with the current sheet at constant 
 tilt) was used in the calculations, with a solar dipole pointing south and a plasma sheet inclined by  20$^{\circ }$-30$^{\circ }$ as observed in 2007.

The initial conditions were defined 
 by assuming that the smaller grains are produced by collisions 
 between larger grains in the interplanetary dust cloud within 1 AU. 
 The initial positions were distributed over the region of the circumsolar 
 dust cloud. The charge-to-mass ratio of the nanoparticles was assumed 
 to be constant, with a set of different values of $q/m$ (from $10^{-6}e/m_p$ to $10^{-4}e/m_p$). The Lorentz  force, gravity, and Poynting-Robertson forces  were taken into account  in the equations of motion. The production rate of nanoparticles, 
 estimated using the collision model and the model of the dust 
 distribution described in Mann and Czechowski (2005), was used to 
 define the probability weight for each trajectory in the sample.

 This simulation shows that particles of $q/m\geq 10^{-5}e/m_p$, i.e. mass $m\leq 10^{-20}$ kg,  move outward and in the prograde direction (that of the solar rotation which winds the magnetic field lines) at about 300 km s$^{-1}$  near  the ecliptic at 1 AU. The speed decreases for larger masses: grains with $q/m= 10^{-6}e/m_p$  are still accelerated, but reach a lower asymptotic speed (150-200 km s$^{-1}$); those with $q/m= 10^{-7}e/m_p$ (of Larmor frequency less than the orbital frequency) do not attain high velocity, but  their initial orbits are significantly disturbed. The simulations  also show that the grains released
within some critical distance from the Sun (0.15 AU), where the magnetic field is very close to radial, do not escape to large distances even if their $q/m$ is high.

The high speed of the grains increases dramatically the electric signal produced upon impact. Dust impact ionisation involves complex processes  not fully included in laboratory simulations and theory at present, so that  the impact generated charge $Q$ is   determined semi-empirically with a large uncertainty (Kissel  and  Kr\"{u}ger, 1987; Hornung  and Kissel, 1994). It varies with the grain's mass and speed typically as
\begin{eqnarray}
Q \simeq 0.7 m^{1.02}v^{3.48}\text{ ,}  \label{Q}
\end{eqnarray}
with $Q$ in Cb, $m$ in kg, $v$ in km s$^{-1}$ \cite{mcb99}. The three coefficients in  (\ref{Q}) depend on mass, speed,  angle of incidence, as well as  grain and target composition, so that $Q$ may differ from this relationship by one order of magnitude.

An extrapolation of   (\ref{Q}) suggests that a nanoparticle impacting at 300 km s$^{-1}$ is equivalent (in terms of  released charge $Q$) to a  grain of mass greater by 4 orders of magnitude impacting at   20 km s$^{-1}$. However, firstly, a speed of 300 km s$^{-1}$ - corresponding to an energy per atom of more than 10 keV (for atomic mass $A \simeq 20$)  should produce strong  effects outside the range of present simulations, and secondly, the grains are much smaller than those considered in the simulations. Therefore, one must verify that this extrapolation satisfies the constraint of energy conservation. Indeed, the  released charge, which in that case comes essentially from the  target, cannot exceed $Q_{\rm{max}} \simeq mv^2/2E_0$, where $E_0$ (in eV) is the minimum energy required to create a free ion or electron out of a solid lattice; the equality would correspond to the case when the entire energy of the impact is spent into ionisation energy.  With $m\simeq 10^{-20}$ kg, $v \simeq 300 $ km s$^{-1}$, and $E_0 \simeq 10$ eV, we have $Q_{\rm{max}}/m \simeq 4.5 \times 10^{9}$ Cb/kg. The charge  (\ref{Q}) is smaller  by a factor of about 40 than this upper limit, and thus amounts to an ionisation efficiency of $2.5 \times 10^{-2}$.

\section{Results and discussion}

The observed values   given in Section~2  then enable us to infer dust properties.  We make two assumptions: first we assume that  $Q$ is given by (\ref{Q}) with $v$ calculated in Section~4, so that  $Q\sim 1.2 \times 10^8 m$ for  $m\leq 10^{-20}$ kg and  varies weakly with $m$ for larger masses because of the decrease in speed shown by the simulation; second we assume that the cumulative dust flux is of the form
\begin{eqnarray}
F= F_0 m^{-\delta }  \label{F}
\end{eqnarray}
(in SI units) with $\delta \simeq 5/6$ according to the interplanetary model in this range, so that the differential flux is $\mid  dF/dm \mid  =\delta F_0 m^{-(\delta +1)}$ and  the differential impact rate contributing to the voltage power spectrum is
\begin{eqnarray}
\mid dN/dm \mid \sim R_2^2 F_0 \delta   m^{-(\delta +1)}\text{ .}  \label{dif}
\end{eqnarray}
Using the pulse amplitude  (\ref{V2}), the expressions (\ref{R2}) for  $R_2$ and (\ref{tau2}) for $\tau _2$,  and integrating  the impact rate (\ref{dif}) over mass, we deduce
\begin{eqnarray}
<N \delta V^2/\tau ^2 >^{1/2} \sim 0.6 \times 10^{11} T_{\rm{eV}} v_{E\rm{(km }\; \rm{ s}^{-1})} F_0^{1/2}  m_{\rm{min}(10^{-20} \rm{kg})}^{-1/12}\text{ ,}  \label{Ndelta}
\end{eqnarray}
where we have used $\tau \sim \tau_2$ with $n_a \sim 5 \times 10^6$ m$^{-3}$, $v_{E\rm{(km \; s}^{-1})}$ is the expansion speed in km s$^{-1}$, and $m_{\rm{min}(10^{-20} \rm{kg})}$ is the minimum mass (in units of  $10^{-20}$ kg) taken as a lower bound of the integral over mass. The result depends very weakly on this minimum mass because of the small exponent $(1/12)$ in (\ref{Ndelta}); it is not much affected either by particles of mass $m>10^{-20}$ kg because the decrease in flux is not compensated by an increase in $Q$.

Comparing (\ref{Ndelta}) with the measurement (\ref{meas}), and assuming $m_{\rm{min}(10^{-20} \rm{kg})} \sim 10^{-2}$, we deduce $F_0$, whence the cumulative flux
\begin{eqnarray}
F \sim 2 \times 10^3   m_{10^{-20} \rm{kg}}^{-5/6} /( T_{\rm{eV}}   v_{E\rm{(km \; s}^{-1})})^2  \text{ m}^{-2}\text{s}^{-1}\text{ ,} \label{FF}
\end{eqnarray}
where $ m_{10^{-20} \rm{kg}}$ is the mass in units of $10^{-20}$ kg. With $T\sim 1-5$ eV and $v_E \sim 20 - 50$ km s$^{-1}$  \cite{hor94}, we find $F \sim [0.03 - 5]   m_{10^{-20} \rm{kg}}^{-5/6}$ m$^{-2}$s$^{-1}$. Note that  $v_E \simeq 20$ km  s$^{-1}$ corresponds to a kinetic energy of the expansion which is equivalent to about 2 eV or larger. The upper value of $v_E$ is determined by the constraints of the observed rise time  and energy conservation.

The above result does not take into account the uncertainty in expression (\ref{Q}) for $Q$, which amounts to a factor of $\sim 10$, given the extrapolation made and the diversity of the spacecraft surfaces (ITO-coated silver teflon perforated to the silver backing material, kapton, solar panel mylar, and a few metallic surfaces  \cite{dri08}). Since $dN/dm \propto R_2^2 \propto Q^{2/3}$, this translates into  an uncertainty in $F$ of a factor $10^{2/3} \simeq 5$. Furthermore, the mass of $10^{-20}$ kg of the detected particles is expected be uncertain by a factor of $\sim 10$ due to uncertainties in the grain mass density, electric potential, and speed. Finally,  since these detections are made during roughly 10\% of the STEREO A observing  time, the average flux is roughly ten times smaller than the value given in (\ref{FF}).

 \begin{figure}    
   \centerline{\includegraphics[width=1\textwidth,clip=]{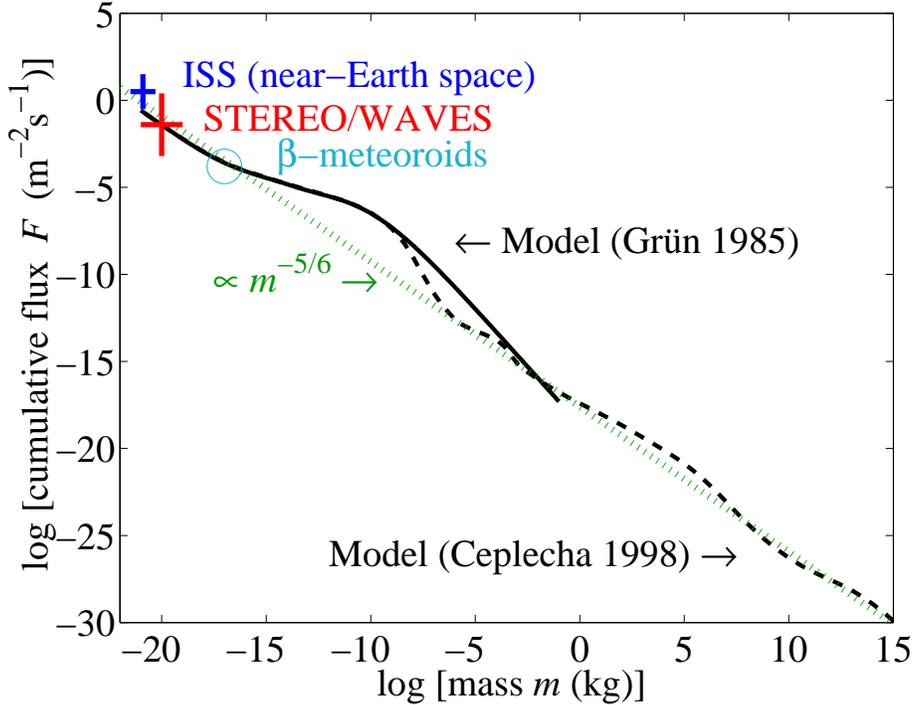}
              }
              \caption{Flux of particles of mass greater than $m$. Our result, the ISS detection (Carpenter   \textit{et al.}, 2007), and the $\beta $ meteoroids detected by Ulysses (Wehry and Mann, 1999) are superimposed to the interplanetary dust flux model (solid line, Gr\"{u}n   \textit{et al.} 1985) and to the model derived from meteor and small solar system object observations (dashed, Ceplecha  \textit{et al.}, 1998). The green dotted line is a flux  $\propto m^{-5/6}$, as expected for collisional fragmentation equilibrium (adapted from Meyer-Vernet, 2007).
        \protect  \label{fig5}   }      
   \end{figure}

In Figure~5  we have superimposed this result  on a composite  model  of interplanetary dust and small bodies (Gr\"{u}n et al., 1985; Ceplecha et al., 1998). Within the relatively large uncertainties, our result is consistent with the model. These newly found particles are difficult to detect with conventional dust detectors because they are outside their calibration range. However,  a   similar  flux (see Figure~5) has been  recently inferred in low Earth orbit on the International Space Station \cite{car07}. Nanoparticles have also been inferred at large distances from comet Halley  \cite{utt90}. Furthermore, streams of fast nanoparticles ejected from the surrounding of Jupiter  \cite{zoo96} and Saturn  \cite{kem05} have been inferred, using conventional dust detectors  below their calibration range.

Following  the  observations described in the present paper, we have examined in detail the wave data on other spacecraft, even though the nanoparticle signals are expected to be small, as discussed in Section 3. On Cassini we have observed    small  voltage pulses  near Jupiter that are consistent with nanoparticle impacts according to Equation (\ref{V1}), when the RPWS frequency receiver was connected in monopole mode. The signal strength was estimated  assuming dust properties inferred from  simultaneous observations with  conventional dust detectors \cite{mey08}. Similar observations  in interplanetary space at various heliocentric distances are now under study.

It is noteworthy  that different S/WAVES  dust observations were  recently reported, which have been analysed during periods when we do not detect nanoparticles. They have been shown to coincide with perturbations of the SECCHI instrument. These voltage pulses, detected with the  time domain sampler, have an impact rate far too small to be observed with the  frequency receiver,   contrary to those reported in the present paper, and correspond to impacts of large grains of radius several microns \cite{stc09}.

The present results are limited for several reasons. The charge produced by impacts of fast nanoparticles is only estimated via extrapolations of experimental results valid for larger particles and smaller velocities. It is essential that future laboratory and numerical simulations of dust impact ionisation include fast nanoparticles, and consider the size range intermediate between bulk materials and heavy ions. Furthermore, we cannot expect to have a clear description of the charge produced by impacts without knowing the composition of the nanoparticles. In addition, we should take into account more correctly the behaviour of the impact plasma in the complex environment of the spacecraft, and perform  more detailed simulations of the response of the wave instrument to impulses.  Two  important questions have to be studied: the variation of the signal with time and longitude, and the sites of impacts, which depend on the direction of the impact velocities. This problem is complex because the solar wind state should affect both the dynamics of the particles  (via the solar wind velocity and magnetic field  along the particle trajectory) and their detection by the electric antennas (via the local solar wind density). In particular, STEREO might also detect fast nanoparticles originating from other sources than the inner solar system.

Finally, it is noteworthy that our result, which is compatible with the model of interplanetary dust  at 1 AU (c.f. Figure~\ref{fig5}), lies on a curve   $\propto m^{-5/6}$ that spans 35 orders of magnitude in mass (green dotted line on  Figure~\ref{fig5}). This curve corresponds to a cascade of destruction and production by collisional fragmentation of objects with a cross-sectional area  $\propto m^{2/3}$ \cite{doh69}.  This agreement in order of magnitude supports the idea that the particles are indeed generated by collisional fragmentation from larger dust particles. The superimposed deviations to the slope are produced by other sinks and sources of particles, in particular ejection of nanoparticles.

\begin{acks}

We thank the team who conceived, designed, built, and tested the instrument, and are very grateful to P.-L. Astier for simulations of the response  to transient pulses. The French part was supported by CNES and  CNRS. A.C. was supported by the Polish Ministry of Science grant 4 T12E 002 30. We thank the International Space Science Institute for fostering discussions within the working group `Dust-plasma interactions: observations in the interplanetary medium and in the environment of solar system objects'. We are grateful to an anonymous referee for helpful comments.

\end{acks}

\end{article} 

\end{document}